\documentclass[epj]{svjour}
\usepackage{graphics}
\begin{document}
\title{Manifestations of anomalous glue: 
light-mass exotic mesons and $g_{\eta' NN}$}
\author{Steven D. Bass \inst{1} %\and Second author\inst{2}% etc
}                     % Do not remove
\institute{High Energy Physics Group, Institute for Experimental Physics 
and Institute for Theoretical Physics, 
Universit\"at Innsbruck, Technikerstrasse 25, A 6020 Innsbruck, Austria}
\date{Received: date / Revised version: date}
\abstract{
The light-mass exotics with $J^{PC}=1^{-+}$ observed at BNL and CERN 
may have a simple explanation as dynamically generated resonances in 
$\eta' \pi$ rescattering in the final state interaction. This dynamics 
is mediated by the same anomalous glue which also generates the large 
mass of the $\eta'$.
OZI violating processes are also potentially important to 
$\eta'$ production in proton-proton collisions close to threshold. 
\PACS{
{12.39.Mk}{Glueball and nonstandard multi-quark/gluon states}
\and {13.75.Lb}{Meson-meson interactions}
\and {13.75.Cs}{Nucleon-nucleon interactions}
     } % end of PACS codes
} %end of abstract
\maketitle
\section{Introduction}
\label{intro}

Searching for evidence of gluonic degrees of freedom in low-energy QCD 
is one of the main themes driving present experimental and theoretical 
studies of the strong interaction. 
Key probes include glueball and $J^{PC}=1^{-+}$ exotic meson 
searches plus OZI violation in $\eta'$ physics \cite{uppsala}
which is sensitive to gluonic degrees of freedom through the 
U(1) axial anomaly \cite{zuoz}.
Exotic mesons are particularly interesting because the quantum numbers 
$J^{PC}=1^{-+}$ are inconsistent with a simple quark-antiquark bound 
state \cite{dzierba}.
Two such exotics, with masses 1400 and 1600 MeV, have been observed 
in experiments at BNL \cite{exoticb} and CERN \cite{exoticc} 
in decays to $\eta' \pi$ and $\eta \pi$.
These exotics have been a puzzle to theorists and experimentalists 
alike because the lightest mass $q {\bar q} g$ state with exotic 
quantum numbers predicted by lattice calculations \cite{lattice,thomas}
and QCD inspired models \cite{models} has mass about 1800-1900 MeV.
As we explain here, the presently observed light-mass exotics seen at 
BNL and CERN may have a simple explanation \cite{bassem} 
as dynamically generated resonances in $\eta' \pi$ rescattering 
(in the final state interaction).
The anomalous glue which generates the large $\eta'$ mass plays an essential 
role in this dynamics.

The physics of anomalous glue also yields interesting phenomenology in the 
$\eta'$-nucleon system.
The flavour-singlet Goldberger-Treiman relation \cite{venez} relates the 
flavour-singlet axial-charge $g_A^{(0)}$ extracted from polarized deep 
inelastic scattering \cite{bass99} to the $\eta'$-nucleon coupling 
constant.
The small value of $g_A^{(0)}$ (about 50\% of the OZI value 0.6) 
measured in polarized DIS and the large mass of the $\eta'$ point 
to large violations of OZI in the flavour-singlet 
$J^{PC} = 1^{++}$ channel.
OZI violating processes may also play an important role \cite{sb99}
in $\eta'$ production in proton-nucleon collisions close to threshold 
\cite{goran}. 
This process is presently under investigation at COSY \cite{cosy}.

We first review the puzzle of light-mass exotics.
We then explain how the presently observed states may be understood as 
dynamically generated resonances in $\eta' \pi$ rescattering.
Here we briefly review the $U_A$(1)-extended effective Lagrangian 
for low-energy QCD \cite{vecca,veccb}. This Lagrangian is constructed 
so that it successfully includes the effects of the strong U(1) axial 
anomaly and the large $\eta'$ mass.
Finally, we discuss $\eta'$ production in proton-nucleon collisions 
close to threshold where new effects \cite{sb99} of 
OZI violation are suggested by coupling this Lagrangian to the nucleon.

\section{Light-mass exotics}

The $J^{PC}=1^{-+}$ light-mass exotics discovered at BNL 
\cite{exoticb} and CERN \cite{exoticc} were observed in 
decays to $\eta \pi$ and $\eta' \pi$.
Two such exotics, denoted $\pi_1$, have been observed 
through
$\pi^- p \rightarrow \pi_1 p$ at BNL \cite{exoticb}:
with masses 1400 MeV (in decays to $\eta \pi$) 
and 1600 MeV (in decays to $\eta' \pi$ and $\rho \pi$).
The $\pi_1 (1400)$ state has also been observed in ${\bar p} N$
processes by the Crystal Barrel Collaboration at CERN \cite{exoticc}. 
While the exotic quantum numbers $J^{PC}=1^{-+}$
are inconsistent with a quark-antiquark bound
state, they can be generated through a ``valence'' 
gluonic component
-- for example through coupling 
to the operator $[ {\bar q} \gamma_{\mu} q G^{\mu \nu}$ ].
However, the observed exotics are considerably lighter than 
the predictions (about 1800-1900 MeV) of quenched lattice QCD 
\cite{lattice,thomas} and  QCD inspired models \cite{models} 
for the lowest mass $q {\bar q} g$ state with $J^{PC}=1^{-+}$.
These results suggest that, 
perhaps, the ``exotic'' states observed by the experimentalists 
might involve significant meson-meson bound state contributions.  
Furthermore,
the decays of the light mass exotics to $\eta$ or $\eta'$ mesons 
plus a pion suggest a possible connection to axial U(1) dynamics.

This idea has recently been investigated \cite{bassem} in a model of 
final state interaction in $\eta \pi$ and $\eta' \pi$ rescattering 
using the $U_A$(1)-extended chiral Lagrangian \cite{vecca,veccb}, 
coupled channels and the Bethe-Salpeter equation, 
following the approach of the Valencia group \cite{valencia}.

The $U_A(1)$-extended low-energy effective Lagrangian used in these
calculations is:
\begin{eqnarray}
{\cal L}_{\rm m} = 
{F_{\pi}^2 \over 4} 
{\rm Tr}(\partial^{\mu}U \partial_{\mu}U^{\dagger}) 
+
{F_{\pi}^2 \over 4} {\rm Tr} \biggl[ \chi_0 \ ( U + U^{\dagger} ) \biggr]
\nonumber \\
+ 
{1 \over 2} i Q {\rm Tr} \biggl[ \log U - \log U^{\dagger} \biggr]
+ {3 \over {\tilde m}_{\eta_0}^2 F_{0}^2} Q^2 
\nonumber \\
+ \lambda \ Q^2 \ {\rm Tr} \ \partial_{\mu} U \partial^{\mu} U^{\dagger}
\end{eqnarray}
%where we work to $O(p^2)$ in the meson momentum.
Here
$
U = \exp \ (  i {\phi \over F_{\pi}}  
                  + i \sqrt{2 \over 3} {\eta_0 \over F_0} )
$
is the unitary meson matrix where $\phi = \sum_k \phi_k \lambda_k$
with $\phi_k$ 
denotes the octet of would-be Goldstone bosons 
$(\pi, K, \eta_8)$
associated with 
spontaneous chiral 
$SU(3)_L \otimes SU(3)_R$ breaking, and $\eta_0$ 
is the singlet boson;
$Q$ denotes the topological charge density
($Q = {\alpha_s \over 4 \pi} G {\tilde G}$).
Also,
$\chi_0 = 
{\rm diag} [ m_{\pi}^2, m_{\pi}^2, (2 m_K^2 - m_{\pi}^2 ) ]$
is the quark-mass induced meson mass matrix,
${\tilde m}_{\eta_0}$
is the gluonic induced mass term for the singlet 
boson and $\lambda$ is an OZI violating coupling -- see below.
The pion decay constant $F_{\pi} = 92.4$MeV and 
$F_0$ renormalises 
the flavour-singlet decay constant
$F_{\rm singlet} = F_{\pi}^2/F_0 \sim 100$MeV.

The gluonic potential involving $Q$ is constructed so that the 
effective theory reproduces the QCD axial anomaly \cite{zuoz}
in the divergence of the gauge invariantly renormalized 
axial-vector current. 
This potential also generates the gluonic contribution 
to the $\eta$ and $\eta'$ masses:
$Q$ is treated as a background field with no kinetic term;
it may be eliminated through its equation of motion to yield
\begin{equation}
{1 \over 2} i Q {\rm Tr} \biggl[ \log U - \log U^{\dagger} \biggr]
+ {3 \over {\tilde m}_{\eta_0}^2 F_{0}^2} Q^2 
\
\mapsto \
- {1 \over 2} {\tilde m}_{\eta_0}^2 \eta_0^2 
\end{equation}
The $\eta$--$\eta'$ mass-matrix resulting from Eq.(1) 
gives
$\eta$ and $\eta'$ masses:
\begin{equation}
m^2_{\eta', \eta} = (m_{\rm K}^2 + {\tilde m}_{\eta_0}^2 /2) 
\pm {1 \over 2} 
\sqrt{(2 m_{\rm K}^2 - 2 m_{\pi}^2 - {1 \over 3} {\tilde m}_{\eta_0}^2)^2 
   + {8 \over 9} {\tilde m}_{\eta_0}^4} .
\end{equation}
If the gluonic term ${\tilde m}_{\eta_0}^2$ 
were zero in this expression, one would have
$m_{\eta'} = \sqrt{2 m_{\rm K}^2 - m_{\pi}^2}$ 
and
$m_{\eta} = m_{\pi}$.
Without any extra input from glue, in the OZI limit, the $\eta$ 
would be approximately an isosinglet light-quark state 
(${1 \over \sqrt{2}} | {\bar u} u + {\bar d} d \rangle$)
degenerate with the pion and 
the $\eta'$ would a strange-quark state $| {\bar s} s \rangle$
--- mirroring the isoscalar vector $\omega$ and $\phi$ mesons.
Indeed, in an early paper \cite{weinberg} Weinberg argued that
the mass of the $\eta$ would be less than $\sqrt{3} m_{\pi}$
without any extra U(1) dynamics to further break the axial U(1) symmetry.
The gluonic contribution 
to the $\eta$ and $\eta'$ masses is about 300-400 MeV \cite{uppsala}.

In the model calculations \cite{bassem} of FSI the meson-meson 
(re-)scattering potentials in the Bethe-Salpeter equation were 
derived from the Lagrangian (1).
The OZI violating interaction
$
\lambda \ Q^2 \ {\rm Tr} \ \partial_{\mu} U \partial^{\mu} U^{\dagger}
$
\cite{veccb}
was found to play a key role in the $J^{PC}=1^{-+}$ channel.
A simple estimate for the coupling $\lambda$ can be deduced 
from the decay $\eta' \rightarrow \eta \pi \pi$ 
yielding two possible solutions with different signs.
Especially interesting is the negative sign solution.
When substituted into the Bethe-Salpeter equation
this solution was found to yield a dynamically generated 
p-wave resonance with exotic quantum numbers $J^{PC}=1^{-+}$.
Furthermore, this resonance was found to have mass $\sim 1400$ MeV and 
width $\sim 300$ MeV -- close to the observed exotics.
(The width of the $\pi_1(1400)$ state measured in decays to $\eta \pi$ 
 is $385 \pm 40$MeV; the width of the $\pi_1(1600)$ measured in decays
 to $\eta' \pi$ is $340 \pm 64$MeV.)
The topological charge density mediates the coupling of the 
dynamically generated light-mass exotic to the $\eta \pi$ and 
$\eta' \pi$ channels in these calculations.
For detailed discussion and the amplitudes for the 
individual channels which contribute to this dynamics, see \cite{bassem}.

\section{OZI violation in the $\eta'$--nucleon system}

Going beyond the meson sector, it is interesting to look for evidence of 
OZI violation in the $\eta'$--nucleon system.  Some guidance is provided 
by coupling the $U_A$(1)-extended chiral Lagrangian to the nucleon \cite{sb99}.
Here we find
%One also finds 
a gluon-induced contact interaction in the 
$pp \rightarrow pp \eta'$ reaction close to threshold: 
\begin{equation}
{\cal L}_{\rm contact} =
         - {i \over F_0^2} \ g_{QNN} \ {\tilde m}_{\eta_0}^2 \
           {\cal C} \
           \eta_0 \ 
           \biggl( {\bar p} \gamma_5 p \biggr)  \  \biggl( {\bar p} p \biggr)
\end{equation}
Here $g_{QNN}$ is an OZI violating coupling which measures 
the one particle irreducible coupling of the topological 
charge density $Q$ to the nucleon and ${\cal C}$ is a 
second OZI violating coupling which also features in $\eta'N$ scattering.
The physical interpretation of the contact term (4) 
is a ``short distance'' ($\sim 0.2$fm) interaction 
where glue is excited in the interaction region of
the proton-proton collision and 
then evolves to become an $\eta'$ in the final state.
This gluonic contribution to the cross-section 
for $pp \rightarrow pp \eta'$ 
is extra to the contributions associated with 
meson exchange models \cite{holinde,wilkin}.
There is no reason, a priori, to expect it to be small.

What is the phenomenology of this OZI violating interaction ?

Since glue is flavour-blind the contact interaction (4) 
has the same size in both 
the $pp \rightarrow pp \eta'$ and $pn \rightarrow pn \eta'$ reactions.
CELSIUS \cite{celsius} have measured the ratio
$R_{\eta} 
 = \sigma (pn \rightarrow pn \eta ) / \sigma (pp \rightarrow pp \eta )$
for quasifree $\eta$ 
production from a deuteron target up to 100 MeV above threshold.
They observed that $R_{\eta}$ is approximately energy independent 
$\simeq 6.5$ over the whole energy range --- see Fig.1.
\begin{figure}
\resizebox{0.4\textwidth}{!}{%
  \includegraphics{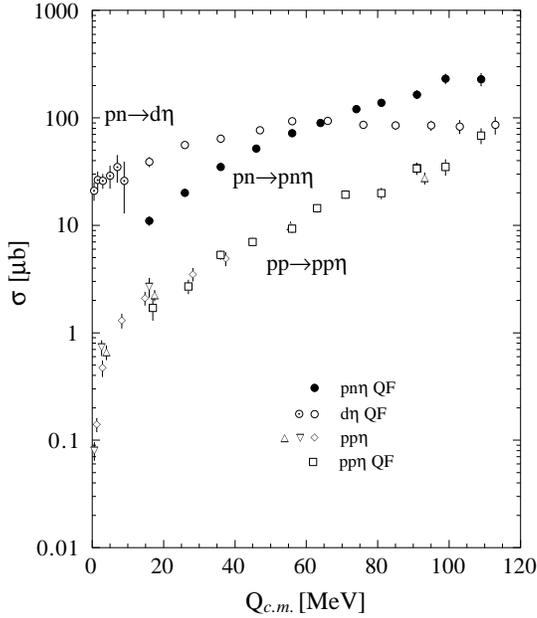}
}
%\vspace{5cm}       % Give the correct figure height in cm
\caption{CELSIUS data on $\eta$ production}
\label{fig:1}       % Give a unique label
\end{figure}
The value of this ratio signifies a strong isovector exchange 
contribution to the $\eta$ production mechanism \cite{celsius}.
This experiment can be repeated for $\eta'$ production.
The cross-section for $pp \rightarrow pp \eta'$ 
close to threshold has been measured at COSY \cite{cosy}.
A new experiment
\cite{cosyprop}
has been initiated 
to carry out the $pn \rightarrow pn \eta'$ measurement.
In the formal limit that the $pp \rightarrow pp \eta'$ 
reaction were dominated by gluonic-induced production,
the ratio
\begin{equation}
R_{\eta'} =
 \sigma (pn \rightarrow pn \eta' ) / \sigma (pp \rightarrow pp \eta' )
\end{equation}
would approach unity close to threshold after we correct for final 
state interaction
\cite{faldt}
between the two outgoing nucleons.
It will be interesting to compare future measurements of
$R_{\eta'}$ with the CELSIUS measurement \cite{celsius}
of $R_{\eta}$.
Given that $\eta'$ phenomenology is characterised 
by large OZI violations, it is natural to expect 
large OZI effects in the $pp \rightarrow pp \eta'$ process.

\vspace{0.5cm}

{\bf Acknowledgements} \\

SDB is supported by a Lise-Meitner Fellowship, M683, of the Austrian FWF.
I thank W. Oelert and the COSY-11 Collaboration for hospitality in 
Juelich, P. Moskal for helpful discussions and E. Marco for collaboration 
on light-mass exotics.


\end{document}